\documentclass[aps,prl,twocolumn,showpacs,floatfix,superscriptaddress]{revtex4}
\usepackage{graphicx}
\usepackage{graphics}
\usepackage{epsfig}
\usepackage{amsmath}
\usepackage{amssymb}
\usepackage{dcolumn}
\usepackage{bm}
\usepackage{longtable}

\begin{document}

\title{Coulomb versus spin-orbit interaction in few-electron 
carbon-nanotube quantum dots}

\author{Andrea Secchi}

\affiliation{Dipartimento di Fisica, Universit\`a 
degli Studi di Modena e Reggio Emilia, Via Campi 213/A, 41100 Modena, Italy}

\affiliation{CNR-INFM National Research Center S3, Via Campi 213/A, 
41100 Modena, Italy}

\author{Massimo Rontani}

\affiliation{CNR-INFM National Research Center S3, Via Campi 213/A, 
41100 Modena, Italy}

\date{\today }

\begin{abstract}
Few-electron states in carbon-nanotube 
quantum dots are studied by means of the configuration-interaction method.
The peculiar non-interacting feature of the tunneling
spectrum for two electrons, recently measured by
Kuemmeth \emph{et al.} [Nature
{\bf 452,} 448 (2008)], is explained by the splitting of a low-lying
isospin multiplet due to spin-orbit interaction.
Nevertheless, the strongly-interacting ground state forms a
``Wigner molecule'' made of electrons localized in space.
Signatures of the electron molecule may be seen in tunneling spectra
by varying the tunable dot confinement potential. 
\end{abstract}


\maketitle 

After almost two decades of research, carbon nanotubes
(CNs) \cite{Saito98} still provide a venue for the 
investigation of fundamental
properties of interacting electron systems, such as Luttinger-liquid
\cite{Bockrath99} and Wigner-crystal \cite{Deshpande08} behavior,
Mott state \cite{Deshpande09},
Kondo effect \cite{Nygard00} and
Andreev transport
\cite{vanDam06} in CN quantum dots (QDs). 
With respect to semiconductor QDs \cite{Reimann02,Kalliakos08},
low-screening, ultra-clean CN QDs appear to be ideal 
candidates \cite{Deshpande08}
for the realization of long-sought ``Wigner molecules'' (WMs)
\cite{Reimann02} of strongly correlated electrons. 
These classical geometrical configurations of
electrons localized in space are insensitive to
the spin state of the system \cite{Deshpande08,Reimann02}.
On the other hand, recently
Kuemmeth {\it et al.}~\cite{Kuemmeth08} showed that 
orbital and spin degrees 
of freedom are entangled by strong spin-orbit interaction
in single-wall CN QDs, disproving
the popular shell model based on the fourfold
degeneracy of QD energy levels \cite{disproved,Oreg00}.
Intriguingly, the
two-electron tunneling spectrum measured in \cite{Kuemmeth08}
was well explained by a non-interacting 
model with spin-orbit coupling. 
So far, it is unclear if the WM state \cite{Reimann02}
may survive to the perturbation induced by 
spin-orbit interaction. 

In this Letter we show that non-interacting
features of the tunneling spectrum, due to spin-orbit coupling,
coexist with the strongly interacting nature of few-electron states,
as seen from configuration interaction (CI, also known as 
exact diagonalization) calculations.
Electrons in realistic dots form one-dimensional
WMs, which may already have been observed in 
experiments \cite{Kuemmeth08,Deshpande08}. 
We predict that molecular signatures appear in 
the excitation spectrum by varying the QD confinement potential. 

We focus on a QD embedded in a semiconducting CN whose length
scale, $\ell_{\text{QD}}$, is smaller that the CN length. Hence,
$\ell_{\text{QD}}$ is the relevant single-particle (SP) length and 
the effects of the CN boundaries may be neglected.
With respect to previous calculations \cite{Egger97,Oreg00,Grifoni08,Bulaev08},
we assume the QD to be defined by an external gate potential, slowly
varying on the lattice scale, which we model as a one-dimensional 
harmonic oscillator (HO) of frequency $\omega_0$. 
The quadratic potential is the low-energy generic form for a soft confinement  
\cite{harmonic}, setting $\ell_{\text{QD}}=(\hbar/m^*\omega_0)^{1/2}$,
where $m^*=\hbar^2/3R\gamma$ is the effective mass, $R$ is
the CN radius, and $\gamma=$ 0.54 eV$\cdot$nm is the graphene
$\pi$-band parameter. 
SP states $\psi_{n\tau_z}\!(\bm{r})= {\cal{N}}F_n(x)
\phi_{\tau_z}\!(\bm{r})$ are obtained
by the envelope-function modulation $F_n(x)$ of bulk states
$\phi_{\tau_z}\!(\bm{r})$ at the two non-equivalent
minima of the 
lowest conduction band. The isospin index $\tau_z=+1$ ($-1$) labels
valley K  
(K$^{\prime}$), $F_n(x)$ is the wave function of the
$n$th HO excited state, and ${\cal{N}}$ is a normalization factor. 
Here $\phi_{\tau_z}\!(\bm{r}) =
\exp(-iy\tau_z/3R)[\psi_{X,A}(\bm{r})+\tau_z\psi_{X,B}(\bm{r})]$,
where $\psi_{X,A}(\bm{r})$ and $\psi_{X,B}(\bm{r})$
are the Bloch tight-binding states for sublattices
$A$ and $B$, respectively, at 
point $X=$ K ($X=$ K$^{\prime}$) in the reciprocal
space for $\tau_z=+1$ ($-1$).   
The isospin $\tau_z=+1$ ($-1$) points to the
(anti)clockwise rotation along
the circumference coordinate $y$, perpendicular to the tube axis $x$.
The interacting Hamiltonian is 
$\hat{H}=\hat{H}_{\text{SP}}+\hat{V}_{\text{FW}}+\hat{V}_{\text{BW}}$,
which includes the SP term $\hat{H}_{\text{SP}}=\sum_{n\tau_z\sigma_z}
\varepsilon_{n\tau_z\sigma_z}
\hat{c}^{\dagger}_{n\tau_z\sigma_z} \hat{c}_{n\tau_z\sigma_z}$, as well as
the two-body terms for forward (FW) and backward (BW)
Coulomb scattering processes
$\hat{V}_{\text{FW}}$ and $\hat{V}_{\text{BW}}$, respectively
\cite{Egger97,Ando06,Grifoni08}.
Here $\hat{c}_{n\tau_z\sigma_z}$ destroys an electron 
occupying the SP orbital $\psi_{n\tau_z}\!(\bm{r})$ with spin $\sigma_z$.
The SP energy $\varepsilon_{n\tau_z\sigma_z}$
includes the dominant term of
spin-orbit interaction due to the CN curvature
\cite{Ando00,Brataas06,Kuemmeth08} as well as 
the contribution
of an axial magnetic field $B$ \cite{Ando00}:  
\begin{equation}
\varepsilon_{n\tau_z\sigma_z} = \varepsilon_n^{\text{HO}} + 
 \Delta_{\text{SO}} \frac{\gamma}{R} \tau_z \sigma_z 
+ \mu_{\text{B}}B \big( \frac{g^*}{2}\sigma_z - 
\frac{m R \gamma}{\hbar^2} \tau_z \big) .
\label{eq:EB}
\end{equation}
Here
$\varepsilon_n^{\text{HO}}=\gamma/3R +\hbar\omega_0(n + 1/2)$,
$\Delta_{\text{SO}}$ is the spin-orbit coupling term,
$\mu_{\text{B}}$ is the Bohr magneton, $g^*$ is the giromagnetic factor.
The two-body terms $\hat{V}_{\text{FW}}$
and $\hat{V}_{\text{BW}}$ are of the type $\hat{V} = 1/2 \sum 
V_{\alpha\beta\gamma\delta}\,\hat{c}^{\dagger}_{\alpha\sigma_z}
\hat{c}^{\dagger}_{\beta\sigma_z^{\prime}}\hat{c}_{\gamma\sigma_z^{\prime}}
\hat{c}_{\delta\sigma_z}$, where $\alpha\equiv(n,\tau_z)$ and
$V_{\alpha\beta\gamma\delta}$ is the matrix element
of the Ohno potential $V(\bm{r}-\bm{r^{\prime}})=
U_0 (1+\epsilon^2\left|\bm{r}-\bm{r^{\prime}}\right|^2U_0^2/e^4)^{-1/2}$,
which interpolates the two limits of Coulomb-like long-range and Hubbard-like
short-range interactions ($\epsilon$ is the relative dielectric
constant, and $U_0=$ 15 eV) \cite{Grifoni08}.
In the envelope function approach \cite{Kohn} the underlying graphene physics
is buried into the precise form of $V_{\alpha\beta\gamma\delta}$'s,
which depend on both Bloch states $\phi_{\tau_z}\!(\bm{r})$ and
envelopes $F_n(x)$. We evaluate $V_{\alpha\beta\gamma\delta}$ 
by considering explicitly the tight-binding expansion of 
$\phi_{\tau_z}\!(\bm{r})$ 
(cf.~\cite{Grifoni08}). FW and BW terms
correspond to direct [$(\tau_z,\tau_z^{\prime})
\rightarrow (\tau_z,\tau_z^{\prime})$] and exchange 
[$(\tau_z,\tau_z^{\prime})
\rightarrow (\tau^{\prime}_z,\tau_z)$, $\tau_z\neq \tau_z^{\prime}$]
isospin scattering
processes, respectively \cite{Egger97,Ando06,Grifoni08}. 

The few-body problem is solved by means of the 
CI method \cite{Reimann02,Rontani06}. We diagonalize 
$\hat{H}$, which is a matrix
in the basis of the Slater determinants $\left|\Phi_i\right>$ obtained by
filling with $N$ electrons in all possible ways the thirty
lowest-energy SP orbitals $\psi_{\alpha}(\bm{r})$. 
We obtain energies and wave functions
of the many-body ground- and excited-states $\left|\Psi^{(n)}\right>$, written
as linear combinations of $\left|\Phi_i\right>$'s,
$\left|\Psi^{(n)}\right>=\sum c_i^{(n)}\left|\Phi_i\right>$,
in each sector of the Fock space labeled by $N$, the $z$-component
of the total spin $S_z$, and the total parity under spatial 
inversion $x\rightarrow -x$.

\begin{figure}
%\vspace{4mm}
\setlength{\unitlength}{1 cm}
\begin{picture}(8.5,5.0)
\put(0.3,-11.3){\epsfig{file=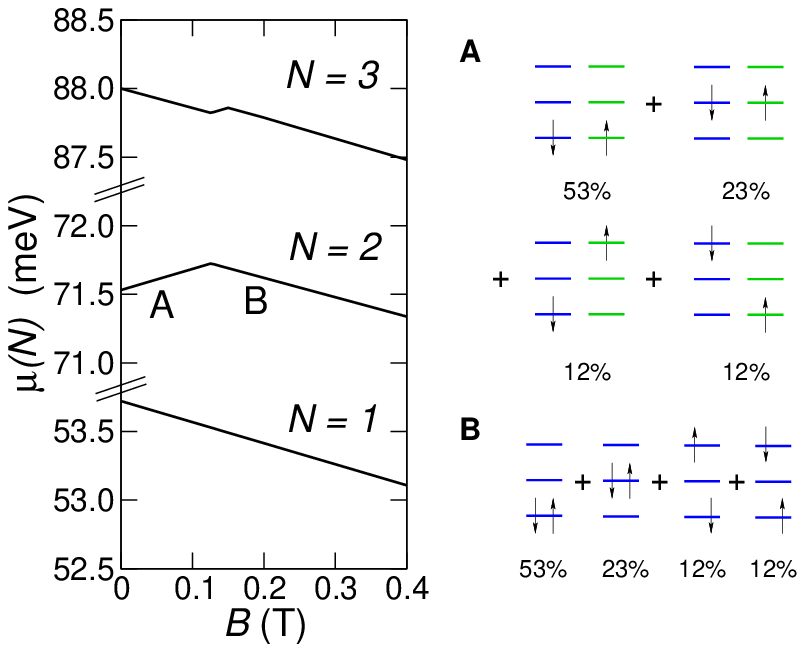,angle=0,width=7.7cm}}
\end{picture}
%\centerline{\epsfig{file=muconf.ps,width= 4cm,angle=0}}
\caption{(color online)  Left: $\mu(N)$ vs $B$
for $1\le N \le 3$. 
Right: Main configurations contributing to states
A and B. Blue (green) levels are HO states 
with isospin $\tau_z=+1$ ($-1$), whereas up- (down-) arrows stand for electron
spin $\sigma_z=+1$ ($-1$). 
The percentages are the weights of the Slater determinants
in the CI expansion of the wave function.
\label{ChemPotentials_Configurations}}
\end{figure}

To compare with tunneling 
spectra \cite{Kuemmeth08}, we compute the chemical potential
$\mu(N)$ for a given value of $B$, 
$\mu_{(i)}\!(N)  = E_{0(i)}(N) - E_0(N-1)$,
where $E_0(N)$ ($E_i(N)$) is the energy of the 
many-body ground ($i$th excited) 
state with $N$ electrons. The predicted
$\mu(N)$'s
may be converted into the gate voltages at which 
electrons tunnel into the QD  
\cite{Reimann02}. We infer from 
\cite{Kuemmeth08} the inputs of CI calculation, i.e.,
$\hbar\omega_0=$ 8 meV, $R=$ 3.6 nm, 
$\Delta_{\text{SO}}=1.24\cdot 10^{-3}$, $g^*=$ 2.14.
The unknown value of $\epsilon$ is a fit parameter 
for a small-gap semiconductor which may be strongly
affected by the leads.
By choosing $\epsilon=3.5$ we obtain the curves of 
Fig.~\ref{ChemPotentials_Configurations}, 
$\mu(N)$ vs $B$ for $1\le N \le 3$,
which compare well with those of Fig.~3(a) in \cite{Kuemmeth08}. 
The plot quantitatively reproduces the dependence
of $\mu(N)$ on $B$, specifically the kink of $\mu(2)$ at
$B_c \approx 0.125$ T and $\mu(3)$ at $\approx 0.15$ T
\cite{disclaimer}.
As a check of consistency with the experiment,
we remark that the CI 
spacing between $\mu(2)$ and $\mu(1)$ at $B=0$, 17.8 meV, 
agrees within 6\% with the value estimated in 
\cite{Kuemmeth08}.

\begin{figure}
%\vspace{4mm}
\centerline{\epsfig{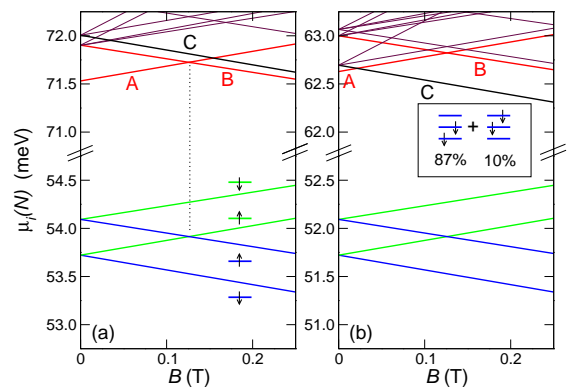}}
\caption{(color online) $\mu_i(N)$ vs $B$ for $N=1,2$, 
for (a) $\hbar \omega_0 = 8$ meV  
and (b) $\hbar \omega_0 = 4$ meV.
Inset: main configurations in the CI expansion of the
wave function of 
state C.
\label{4meV_1_2_chemPot}}
\end{figure}

The kink of $\mu(2)$ at $B_c$
is due to the crossing between different ground
states of two electrons, labeled A (B) for $B<B_c$ ($B>B_c$) 
[cf.~Fig.~\ref{ChemPotentials_Configurations}].
To gain insight into their nature, let us briefly
consider also excited-state contributions to $\mu_i(N)$,
which are shown in Fig.~\ref{4meV_1_2_chemPot}(a)
for $N=1,2$. Note that  Fig.~\ref{4meV_1_2_chemPot}(a)
perfectly matches Fig.~3(c) of \cite{Kuemmeth08}.
The four lowest-energy SP levels $\mu_i(1)$
are explicitly shown. Blue (green) lines indicate states with
$\tau_z=1$ ($-1$), whose orbital 
magnetic moment is (anti)parallel to the CN axis, decreasing 
(increasing) its energy with $B$. The fourfold
degeneracy of the levels ($\tau_z=\pm 1$, $\sigma_z=\pm 1$)
at $B=0$ is split by spin-orbit interaction, which 
entangles orbital and spin degrees of freedom \cite{Kuemmeth08}.
This induces an isospin transition ($\tau_z=-1\rightarrow \tau_z=1$)
for the first excited non-interacting level 
$\mu_1(1)$ at $B\approx$ 0.125 T.
Remarkably, this field is exactly the \emph{same}
as the critical value $B_c$ at which the A $\rightarrow$ B transition
occurs [cf.~the vertical
dashed line in  Fig.~\ref{4meV_1_2_chemPot}(a)]. Because
of this feature,
it was argued in \cite{Kuemmeth08} that the A 
$\rightarrow$ B transition may be solely explained in terms
of spin-orbit interaction. 
The state A was supposed to be a
single Slater determinant with the electrons in the two lowest spin-orbitals,
($\tau_z=1$, $\sigma_z=- 1$) and ($\tau_z=-1$, $\sigma_z=1$), whereas
B is obtained by moving the electron from ($\tau_z=-1$, $\sigma_z=1$) to
($\tau_z= 1$, $\sigma_z=1$).  
In this picture correlation effects are absent.
We next show that A and B are instead strongly interacting states.

The right panel of Fig.~\ref{ChemPotentials_Configurations} shows the Slater
determinants with the largest weights in the CI expansion of $\left|A\right>$.
The blue (green) ladders of levels depict the HO states 
for $\tau_z=+1$ ($-1$), whereas
arrows represent spins. The main configuration, 
whose weight is 53\%, 
is the Slater determinant proposed in \cite{Kuemmeth08} for the
ground state. However, CI calculation shows that there are other three
relevant determinants where also the
excited states of the HO are populated
(Fig.~\ref{ChemPotentials_Configurations}).
This is due to the correlated character
of $\left|A\right>$, the strongest the impact of Coulomb interaction,
the largest the mixing of determinants. Similarly, state B shown in
the right panel of Fig.~\ref{ChemPotentials_Configurations} is
correlated as well. Besides,
state B may be obtained from A by replacing
the levels with $\tau_z=-1$ with those
with $\tau_z=1$ (Fig.~\ref{ChemPotentials_Configurations}).
Hence, A and B belong to a \emph{isospin multiplet},
only differing in the projections $T_z$ of the isospin. 
Here the isospin vector is 
defined by $\hat{\bm{T}}=2^{-1}\sum_{n\tau_z\tau_z'\sigma_z}
\hat{c}^{\dagger}_{n\tau_z\sigma_z}\bm{\sigma}_{\tau_z\tau_z'}
\hat{c}_{n\tau_z'\sigma_z}$, where the components of  
$\bm{\sigma}$ are the Pauli matrices [e.g., 
$\hat{T}_z = (\hat{n}_{+1}-\hat{n}_{-1})/2$ with $\hat{n}_{\tau_z}=
\sum_{n \sigma_z}\hat{c}^{\dagger}_{n\tau_z\sigma_z}
\hat{c}_{n\tau_z\sigma_z}$].
%%%%  Figura 1 Max
%\begin{figure}
%\centerline{\epsfig{file=./splittings.eps,width=2.9in,,angle=0}}
%\caption{(color online)  Lowest CI two-electron energies vs
%$\Delta_{\text{SO}}$.
%Red (black) lines correspond to even (odd) parity states.
%Numbers label level degeneracies. The blue curve is the average
%value of the total spin $\langle\hat{S}^2 \rangle$ 
%evaluated on state A.
%\label{switchingSO_2el}}
%\end{figure}
This is confirmed by extrapolating the behavior of A and
B to the limit $\Delta_{\text{SO}}\rightarrow 0$ (not shown).
In this limit,
A and B and other states at higher energies
collapse into a sixfold
multiplet, wich includes three spin triplet ($T=0$, $S=1$,
$S_z=\pm 1,0$) plus three isospin triplet ($S=0$, $T=1$,
$T_z=\pm 1,0$) states. In fact,
the total wave functions must be odd whereas their (unique) orbital part
is even under particle exchange.
Therefore, except for a 
tiny residual splitting of $\approx$ 2 $\mu$eV
due to BW interaction,
the A-B energy separation
depends on spin-orbit interaction only.

From Eq.~(\ref{eq:EB})
and the inspection of CI wave functions of 
Fig.~\ref{ChemPotentials_Configurations}, it is clear that
the energy splitting between A and B at zero field, 
$\mu_1(2)-\mu(2)=2\Delta_{\text{SO}}\gamma/R$, is the
same as that between one-electron levels,
$\mu_1(1)-\mu(1)$ [cf.~Fig.~\ref{4meV_1_2_chemPot}(a)].
Besides, $\mu(N)$ depends on $B$ through the spin and orbital
magnetic dipole moments
[cf.~Eq.~(\ref{eq:EB})],
which are linear in $\Delta S_z
= S_z(N)-S_z(N-1)$ and $\Delta T_z = T_z(N)-T_z(N-1)$, respectively. 
It is immediate to verify that $\Delta S_z$ and $\Delta T_z$ are
the same for both $\mu(2)$ and 
$\mu_1(1)$, for $B<B_c$ ($\Delta S_z = 1/2$, $\Delta T_z = -1/2$)
and $B>B_c$ ($\Delta S_z=1/2$, $\Delta T_z = 1/2$).
This explains why the critical value of the field for 
both $\mu(2)$ and $\mu_1(1)$ is the same despite
the correlated nature of A and B, in agreement
with the key experimental observation of \cite{Kuemmeth08}. 

The non-interacting feature of $\mu(2)$ discussed above
is \emph{not} universal and may be affected by 
electron correlation. This is the case as one changes e.g.~the QD 
potential, which is controlled in the laboratory
by a capacitively coupled gate.
This in turn changes the ratio of Coulomb matrix elements 
$V$ to the SP spacing $\hbar\omega_0$.
Figure \ref{4meV_1_2_chemPot}(b) shows the analogous plot of 
Fig.~\ref{4meV_1_2_chemPot}(a) for half the value of 
the confinement energy, $\hbar\omega_0=$ 4 meV. The pattern
of $\mu(2)$ vs $B$ has now changed with respect to 
Fig.~\ref{4meV_1_2_chemPot}(a),
due to the crossing between A and a new state, labeled C in the plot,
occurring close to the origin, at $B \approx 0.02$ T. This critical
value depends on the splitting $\Delta_{\text{CA}}$ between C and A
at zero field, which in turn is sensitive to Coulomb correlation.

\begin{figure}
%\vspace{4mm}
\setlength{\unitlength}{1 cm}
\begin{picture}(8.5,5.6)
\put(0.5,0){\epsfig{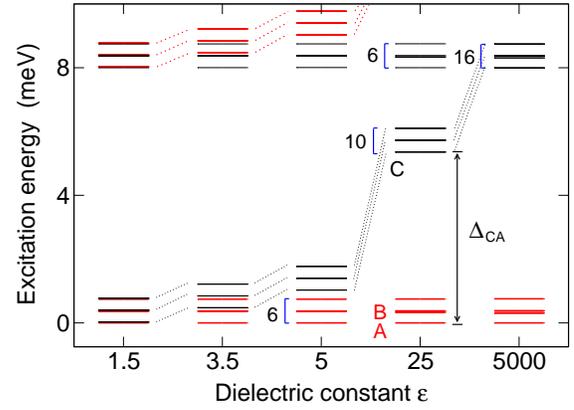}}
\end{picture}
\caption{(color online)  Two-electron excitation energies vs $\epsilon$,
reckoned from the ground state. 
Red (black) lines correspond to even (odd) parity states.
\label{epsilon_depend}}
\end{figure}

The crucial role of Coulomb interaction in the QD is fully appreciated
by considering the excitation spectrum. Figure
\ref{epsilon_depend} shows the two-electron
excitation spectrum vs the
dielectric constant $\epsilon$, which affects
the relevance of correlation effects.  
The lower bound on the horizontal
axis ($\epsilon=1.5$) mimics low screening, typical of 
large-gap semiconducting CNs, whereas the upper bound ($\epsilon=5000$)
may be regarded as the non-interacting limit.
In the latter, all levels bunch into the HO
levels, uniformly spaced by $\hbar\omega_0$ ($=8$ meV). 
In the lowest set of levels (red lines) the two electrons occupy the 
nodeless orbitals in one of the two valleys,
whereas the next set (black lines) is  obtained by promoting one
electron into the first HO excited state.
The states of the first (second) set have even (odd)
parity. Within each set, a residual fine structure survives,
entirely due to spin-orbit interaction. As indicated in 
Fig.~\ref{epsilon_depend}, the total
number of levels in the first (second) set is 6 (16),
given by the possible ways to arrange the two electrons
either in the same or in different valleys compatibly 
with Pauli's exclusion principle.

As $\epsilon$ is reduced in
Fig.~\ref{epsilon_depend}, Coulomb interaction 
alters the energy spectrum. In fact,
the sixteen odd levels belonging to the second set separate
into two multiplets. A first sixfold multiplet is
insensitive to $\epsilon$, whereas a second tenfold multiplet 
experiences a sudden energy drop. 
The former multiplet is associated to the collective motion
of the center of mass (Kohn mode), which is
decoupled from the relative motion and hence 
unaffected by Coulomb interaction \cite{Reimann02}. The sixfold
degeneracy of this odd multiplet, lifted only
by spin-orbit interaction, is the same as that of
the lowest even multiplet, since they
differ only in the excitation of the center of mass coordinate.
The second odd multiplet
is instead sensitive to $\epsilon$, i.e., 
Coulomb interaction. In fact, 
$\Delta_{\text{CA}}$  becomes a small fraction of $\hbar\omega_0$
at the experimental value of $\epsilon=3.5$ ($\Delta_{\text{CA}}/
\hbar\omega_0\approx 0.06$), whereas it
vanishes at $\epsilon = 1.5$. 
As discussed below, such vanishing points to
the formation of a WM, the
state where Coulomb correlation localizes electrons 
in space to minimize their electrostatic energy
\cite{Reimann02,Deshpande08}.

\begin{figure}
%\vspace{4mm}
\setlength{\unitlength}{1 cm}
\begin{picture}(8.5,6.2)
\put(-5.2,-18.0){\epsfig{file=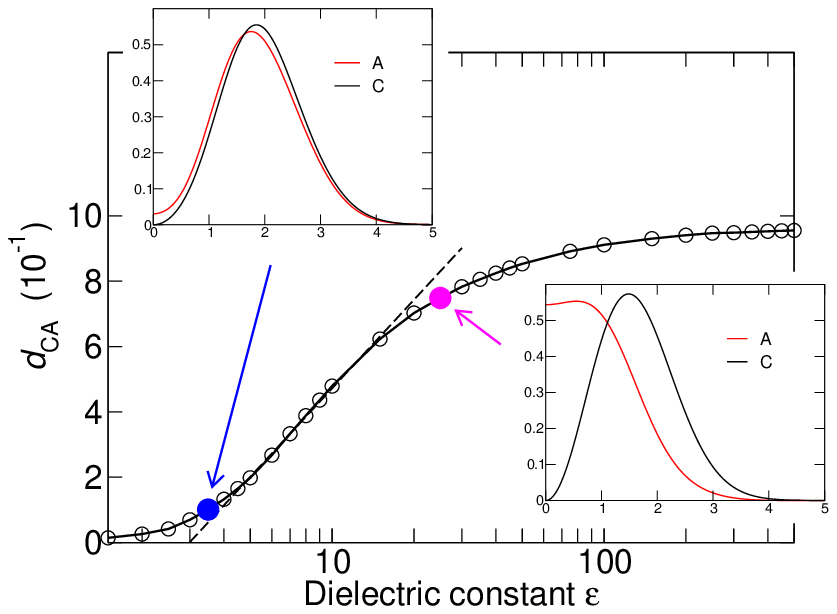, width=20.5cm, angle=0}}
\end{picture}
\caption{(color online)  $d_{\text{CA}}$ vs $\epsilon$. The dashed
line is a linear fit to data in the range $5\le \epsilon \le 15$
with correlation coefficient 0.999.
Insets: $g_{\text{A}}(x)$ and  $g_{\text{C}}(x)$ vs $x$, 
for $\epsilon = 3.5$
(top left) and 25 (bottom right). The length unit is
$\ell_{\text{QD}}$.
\label{WF}}
\end{figure}

The insets of Fig.~\ref{WF} show the correlation function $g(x)$
vs $x$ for states A and C. $g(x)$ is
the probability of finding the two electrons at relative distance
$x=x_1-x_2$ [normalized as $\int_0^{\infty}g(x)dx=1$, with $g(x)=g(-x)$].
As $\epsilon$ decreases from $\epsilon=25$ (bottom right inset) to
the experimental value $\epsilon=3.5$ (top left inset), 
the ground-state probability distribution
$g_{\text{A}}(x)$ (red curve) develops a 
well defined peak at $x_0=1.7$, showing that electrons 
localize in space and freeze their mutual distance.
$x_0$ compares well with the equilibrium 
value $x_{0,\text{cl}}=1.6$ of two point-like
\emph{classical} particles in the HO trap interacting via the
Coulomb potential $e^2/\epsilon\left|x\right|$.
Note that crystallization proceeds by removing probability weight from
$g_{\text{A}}$ at the origin. As localization is fully accomplished, 
$g_{\text{A}}(x=0)\rightarrow 0$,
which is compatible with both even and odd states like A and C, 
respectively. In fact, $g_{\text{A}}(x)$ (red curve) and 
$g_{\text{C}}(x)$ (black curve) tend to coincide as well as  
$\Delta_{\text{CA}} \rightarrow 0$ as $\epsilon$ decreases
(cf.~Fig.~\ref{epsilon_depend}). 
Similarly, as the overlap between the probability weights
of localized electrons is suppressed, exchange interaction is negligible
and different values of $S_z$ ($S_z=0$ for A and $S_z=-1$ for C)
are admissible. In this limit states A and C, which 
only differ for now irrelevant quantum numbers like parity and 
spin, represent the same ``classical'' configuration.

We exploit the progressive overlap between $g_{\text{A}}(x)$
and $g_{\text{C}}(x)$ as $\epsilon$ is reduced to 
characterize the transition to the WM state.
In Fig.~\ref{WF} we plot vs  $\epsilon$ the functional
distance $d_{\text{CA}}$ between states A and C, defined
as $d_{\text{CA}}=\int_0^{\infty}dx\left|g_{\text{A}}(x)
-g_{\text{C}}(x)\right|$.
The semilog plot allows to identify three separate regions,
where $d_{\text{CA}}$ scales differently with $\epsilon$ 
(there are no sharp transitions in finite-size systems).
For $\epsilon > 20$, $d_{\text{CA}}$ slowly tends to the upper bound
1, since the location of the maximum of $g_{\text{A}}(x)$ approaches 
the origin whereas $g_{\text{C}}(x)$ has a node there.
The large probability
of finding two particles close to each other ($x\approx 0$)   
highlights the absence of a correlation hole in the
ground state (cf.~the magenta dot and related plot). 
In the crossover region, $5 < \epsilon < 20$,
$d_{\text{CA}}\propto \log{\epsilon}$
as shown by the linear fit in Fig.~\ref{WF} (dashed line). 
Here a significant correlation hole rapidly forms in A as 
$\epsilon$ decreases. The WM corresponds to
$\epsilon < 5$, where $d_{\text{CA}}  < 0.1$ and it
slowly decreases with $\epsilon$, as $g_{\text{A}}(x)$ and $g_{\text{C}}(x)$ 
overlap almost perfectly. Remarkably, the observed case \cite{Kuemmeth08} 
of $\epsilon=3.5$ (blue dot in Fig.~\ref{WF})
occurs in this region.

The squeezing of the QD confinement potential via
an external gate is a handle to drive Wigner crystallization,
since the effect of lowering $\hbar\omega_0$  
[Fig.~\ref{4meV_1_2_chemPot}(b)] is similar to that of decreasing 
$\epsilon$ (Figs.~\ref{epsilon_depend} and \ref{WF}).
In fact, the critical $B$-value of the
$\text{A}\rightarrow \text{C}$ transition 
reported in Fig.~\ref{4meV_1_2_chemPot}(b) is a measure
of the vanishing of $\Delta_{\text{CA}}$. As $\hbar\omega_0$ 
is reduced, this critical field approaches zero, implying that the
spin-polarized phase C may be induced with no energy cost.
The latter behavior has been observed for hole 
WMs \cite{Deshpande08}.

We have shown that spin-orbit 
and strong Coulomb interaction coexist
in CN QDs, leading to the formation of
WMs at experimentally attainable regimes.
This insight into the
entangled orbital and spin degrees of freedom  
is relevant for the 
all-electrical \cite{Kuemmeth08} and -optical 
\cite{Galland08} manipulation of electron spins in CN-based devices. 
\newline \emph{Note added.} After the submission of this work, Wunsch
reported similar results for a square-well QD \cite{Wunsch09}.  

We thank F. Manghi, E. Molinari, E. Andrei,
G. Steele for stimulating discussions.
This work is supported by INFM-CINECA Supercomputing Project 2008-2009.


\begin{thebibliography}{30}
\bibitem{disclaimer}
The energy conversions of Figs.~2(c) and 3(a) in \cite{Kuemmeth08}
are inconsistent. We take the voltage-to-energy ratio of Fig.~2(c) 
as a reference.
\end{thebibliography}
\end{document}